\documentclass{ws-ijmpb}

\usepackage{graphicx,psfrag,bbm,latexsym,color,dcolumn,bm,dsfont,bbm,color,mathrsfs,bbold}

\newcommand{\ua}{\uparrow}
\newcommand{\da}{\downarrow}
\newcommand{\media}[1]{\left\langle #1 \right\rangle_N}
\newcommand{\E}{\mbox{$\mathsf E$}}
\newcommand{\s}[2]{ #1_{#2}^{} }

\begin{document}

\markboth{C. Presilla and M. Ostilli}{Ground state of many-body lattice systems via a central limit theorem}

\catchline{}{}{}{}{}

\title{Ground state of many-body lattice systems via a central limit theorem}

\author{CARLO PRESILLA}
\address{Dipartimento di Fisica, Universit\`a di Roma ``La Sapienza'',\\
Piazzale A. Moro 2, Roma 00185, Italy\\
Center for Statistical Mechanics and Complexity, 
Istituto Nazionale per la Fisica della Materia, 
Unit\`a di Roma 1, Roma 00185, Italy\\
Istituto Nazionale di Fisica Nucleare, Sezione di Roma 1, Roma 00185, Italy}

\author{MASSIMO OSTILLI}
\address{Dipartimento di Fisica, Universit\`a di Roma ``La Sapienza'',\\
Piazzale A. Moro 2, Roma 00185, Italy\\
Center for Statistical Mechanics and Complexity, 
Istituto Nazionale per la Fisica della Materia, 
Unit\`a di Roma 1, Roma 00185, Italy}


\maketitle

\begin{history}
\received{DAY MONTH YEAR}
\revised{DAY MONTH YEAR}
\end{history}

\begin{abstract}
We review a novel approach to evaluate the ground-state properties
of many-body lattice systems based on an exact probabilistic representation
of the dynamics and its long time approximation via a central limit theorem.
The choice of the asymptotic density probability used in the calculation
is discussed in detail.
\end{abstract}



\section{Introduction}

The real- or imaginary-time dynamics 
of systems described by a finite Hamiltonian matrix, representing
bosonic or fermionic degrees of freedom, admits an exact
probabilistic representation in terms of a proper collection 
of independent Poisson processes \cite{DAJLS,DJS,PRESILLA}.
For a lattice system, the Poisson processes are associated 
to the links of the lattice and the probabilistic representation 
leads to an optimal algorithm \cite{PRESILLA} which coincides with 
the Green Function Quantum Monte Carlo method 
in the limit when the latter becomes exact \cite{SORELLACAPRIOTTI}.

In the recent Ref.~\refcite{OP1} we have exploited the above 
probabilistic representation to derive analytical expressions for the matrix
elements of the evolution operator in the long time limit.
In this way, the ground-state energy as well as the expectation
of a generic operator in the ground state of a lattice system without 
sign problem are obtained as the solution of a simple scalar equation.
The result is based on the application of a central limit theorem to 
the rescaled multiplicities of the values assumed by the potential 
and hopping energies in the configurations dynamically visited 
by the system.
As a consequence, the probabilistic expectations can be calculated by
using a Gaussian-like probability density.
In this paper, we briefly review the approach developed in Ref.~\refcite{OP1} 
and discuss in detail the choice of the asymptotic probability 
density used in the calculation.

\section{Exact probabilistic representation of lattice dynamics}
\label{representation}

We illustrate our approach in the case of imaginary-time dynamics 
for a system of hard-core bosons described by the Hamiltonian
\begin{eqnarray}
\label{Hubbard}
\hat{H} &=& - 
\sum_{i \neq j \in \Lambda} 
\sum_{\sigma=\ua\da} \eta_{ij}
c^\dag_{i\sigma} c^{}_{j\sigma} + \hat{V},
\end{eqnarray}
where $\Lambda\subset Z^d$ is a finite $d$-dimensional lattice
with $|\Lambda|$ sites
and $c_{i \sigma}$ the commuting destruction operators at site $i$ 
and spin index $\sigma$ with the property $c_{i\sigma}^2=0$.
The potential operator $\hat{V}$ is arbitrary, 
\textit{e.g.} for the Hubbard model
$\hat{V} = \sum_{i \in \Lambda} \gamma_i 
c^\dag_{i\ua} c^{}_{i\ua} c^\dag_{i\da} c^{}_{i\da}$.
For simplicity, we assume $\eta_{ij}=\epsilon$  
if $i$ and $j$ are first neighbors and $\eta_{ij}=0$ otherwise.

In order to study the ground-state properties of the Hamiltonian 
$\hat{H}$ it is sufficient to evaluate the long time behavior of 
$\sum_{\bm{n}} \langle \bm{n} |e^{- \hat{H}t} | \bm{n}_0 \rangle$,
where $\bm{n}= (n_{1 \ua},n_{1 \da}, \ldots, n_{|\Lambda| \ua},
n_{|\Lambda| \da})$ are the lattice occupation numbers 
taking the values 0 or 1.
In fact, the ground-state energy is given by
\begin{equation}
\label{E0} 
E_0 = \lim_{t \to \infty}
- \partial_t \log \sum_{\bm{n}} \langle \bm{n}|e^{-\hat{H}t} | \bm{n}_0\rangle,
\end{equation}
while the quantum expectation of a generic operator $\hat{O}$ 
in the ground state of $\hat{H}$ can be obtained via the Hellman-Feynman 
theorem \cite{OP1} by evaluating the 
ground-state energy $E_0(\xi)$ of the modified Hamiltonian 
$\hat{H}+\xi \hat{O}$.

At any finite time $t$, the matrix elements of the evolution operator 
considered above admit the exact probabilistic representation
\begin{equation}
\sum_{\bm{n}} \langle \bm{n} |e^{- \hat{H}t} | \bm{n}_0 \rangle
= \E  \left( \mathcal{M}^t_{\bm{n}_0} \right),
\label{eq1}
\end{equation} 
where $\mathcal{M}^t_{\bm{n}_0}$ is a stochastic functional defined
in terms of independent Poisson processes associated to the links 
of the lattice, see Ref.~\refcite{PRESILLA} for a detailed description.
At each jump of a Poisson process relating sites $i$ and $j$ with 
spin $\sigma$ and taking place at a given configuration $\bm{n}$,
a particle of spin $\sigma$ moves from site $i$ to site $j$
or vice versa if the mod 2 sum 
of the occupations of these two sites is 
$\lambda_{ij\sigma} (\bm{n})= 1$, 
while the lattice configuration $\bm{n}$ remains 
unchanged if $\lambda_{ij\sigma} (\bm{n}) = 0$.
Hereafter, links with $\lambda_{ij\sigma} = 1$ will be called active.
By ordering the jumps according to the times $s_{k}$, $k=1,\dots, N_{t}$, 
at which they take place in the interval $[0,t)$,
we define a trajectory as the Markov chain
$\bm{n}_{1}, \bm{n}_2, \dots, \bm{n}_{N_{t}}$ 
generated from the initial configuration $\bm{n}_0$.
The number of jumps $N_t$ is, of course, a random integer associated to
each trajectory.
We associate to each trajectory also two sequences,
$A_{0},A_{1}, \dots, A_{N_{t}-1}$ and $V_{0},V_{1} \dots, V_{N_{t}}$, 
representing the number of active links and the potential energy 
of the visited configurations
\begin{eqnarray}
A_k &=& \sum_{(i,j) \in \Gamma} ~ \sum_{\sigma=\ua\da} 
\lambda_{ij\sigma}(\bm{n}_k),
\\
V_k &=& \langle \bm{n}_k|\hat{H}| \bm{n}_k \rangle =
\langle \bm{n}_k|\hat{V}| \bm{n}_k \rangle.
\end{eqnarray}
Here, $\Gamma$ is the set of system links, \textit{i.e.} 
the pairs $(i,j)$ with $i<j$ and  
$i,j\in\Lambda$ such that  $\eta_{ij}\neq 0$. 
The stochastic functional $\mathcal{M}^t_{\bm{n}_0}$ which appears
in Eq. (\ref{eq1}) actually depends on the jump times 
$s_1,s_2,\ldots,s_{N_t}$ and on the corresponding sequences 
$A_{0},A_{1}, \dots, A_{N_{t}-1}$ and $V_{0},V_{1} \dots, V_{N_{t}}$.

\section{Probabilistic expectation in the long time limit} 
Evaluating the expectation 
$\E  \left( \mathcal{M}^t_{\bm{n}_0} \right)$ 
over the detailed sequences above specified can be done numerically 
by a Monte Carlo method \cite{PRESILLA}.
In Ref.~\refcite{OP1} we have demonstrated that  
an analytical expression of $\E \left( \mathcal{M}^t_{\bm{n}_0} \right)$ 
can be obtained in the limit of long times.
This result is reached in four steps described in the next subsections.
The crucial point is that, if one integrates over all the
possible jumps times, what matter are not the detailed sequences
$A_{0},A_{1}, \dots, A_{N_{t}-1}$ and $V_{0},V_1 \dots, V_{N_{t}}$ but
the multiplicities $N_A$ and $N_V$ of the possibles values which the 
variables $A$ and $V$ may assume.
We call $\mathscr{A}$ and $\mathscr{V}$ the sets of these values 
and $m_\mathscr{A}$ and $m_\mathscr{V}$ their cardinalities. 
It is clear that the nature of these sets depends only on the structure
of the system Hamiltonian, not on the values of the Hamiltonian parameters.
The expectation $\E \left( \mathcal{M}^t_{\bm{n}_0} \right)$ is 
reduced to an average over $N_A$ and $N_V$.
For $t \to \infty$, this average can be evaluated analytically
by using saddle-point techniques and a central limit theorem.

\subsection{Canonical decomposition of the expectation}
Referring to Ref.~\refcite{OP1} for the details, we 
decompose the expectation as a series of conditional 
expectations with a fixed number of jumps (canonical averages)
\begin{eqnarray}
\label{EXPANSION}
\E \left( \mathcal{M}^t_{\bm{n}_0} \right) &=&
\sum_{N=0}^{\infty} \E \left( \mathcal{M}^t_{\bm{n}_0} |N_{t}=N \right) .
\end{eqnarray}
Integrating over the $N$ jumps times, 
each term of the series (\ref{EXPANSION}) can be written as
\begin{eqnarray}
\label{AVERAGE}
\E \left( \mathcal{M}^t_{\bm{n}_0} |N_{t}=N \right) = 
\media{
\mathcal{W}_N(t)
\prod_{A \in \mathscr{A}} A^{N_A}},
\end{eqnarray}
where $\media{\cdot}$ means average 
over the trajectories with $N$ jumps generated
by extracting with uniform probability one of the active links 
available at the configurations 
$\bm{n}_0,\bm{n}_1,\ldots,\bm{n}_{N-1}$, and $\mathcal{W}_N(t)$, 
named weight, is defined as
\begin{eqnarray}
\label{weights}
\mathcal{W}_N(t) =
\epsilon^{N} \int_{0}^{t}ds_{1} 
\int_{s_{1}}^{t}ds_{2} 
\dots 
\int_{s_{N-1}}^{t}ds_{N}
e^{-V_{0}s_{1}
-V_1(s_{2}-s_{1})
- \dots 
-V_N(t-s_{N})}.
\end{eqnarray}

\subsection{Evaluation of the weights}
According to their definition, the weights satisfy a recursive 
differential equation which is easily solved in terms of the Laplace 
transform
$ \widetilde{\mathcal{W}}_{N}(z) $ \cite{OP1}, \textit{i.e.}
\begin{eqnarray}
\label{WTILDE}
\widetilde{\mathcal{W}}_{N}(z) = 
\epsilon^{N} \prod_{V\in\mathscr{V}} \frac{1}{(z+V)^{N_V}}.
\end{eqnarray}
While this expression shows that $\mathcal{W}_N(t)$ depends
on the multiplicities $N_V$ for any value of $N$, 
the explicit inversion of the Laplace transform can be done
analytically only for $N$ large.
However, this is the limit we are interested in since 
the weights $\mathcal{W}_{N}(t)$ have a maximum at some $N$ which 
increases by increasing $t$. 
By using a complex saddle-point method which is asymptotically exact 
for $N\to \infty$, we get \cite {OP1}
\begin{eqnarray}
\label{WSADDLE}
\mathcal{W}_{N}(t) =
\frac{e^{x_{0}t-\sum_{V\in\mathscr{V}} N_{V} \log[(x_{0}+V)/\epsilon]}}
{\sqrt{ 2\pi \sum_{V\in\mathscr{V}}\frac{\epsilon^2 N_{V}}{(x_{0}+V)^{2}} } },
\end{eqnarray}
where $x_{0}$ is the solution of the equation
\begin{eqnarray}
\label{X}
\sum_{V\in\mathscr{V}}\frac{N_{V}}{x_{0}+V}=t.
\end{eqnarray}

\subsection{Canonical averages via a central limit theorem}
\label{centrallimit}
To evaluate the canonical averages it is useful to introduce
the frequencies, $\s{\nu}{V}=N_V/N$ and $\s{\nu}{A}=N_A/N$, 
which for $N$ large become continuously distributed in the range
$[0,1]$ with the constraints 
\begin{eqnarray}
\label{CONSTRAINTS}
\sum_{V\in\mathscr{V}} \s{\nu}{V} = \sum_{A\in\mathscr{A}} \s{\nu}{A} =1.
\end{eqnarray}
Note that for $N$ large we will not distinguish the different normalizations,
$N+1$ and $N$, of $N_V$ and $N_A$, respectively.
Equation (\ref{AVERAGE}) can be then rewritten as
\begin{eqnarray}
\label{AVERAGE1}
\media{ \mathcal{W}_{N}(t) \! \prod_{A\in \mathscr{A}} A^{N_A} } =
\int d\bm{\nu}
\mathcal{P}_N(\bm{\nu}) 
\frac{e^{x_{0}t + N \left( \bm{\nu},\bm{u} \right) }}
{ \sqrt {2\pi N 
\sum_{V\in\mathscr{V}}\frac{\epsilon^2 \s{\nu}{V}}{(x_{0}+V)^{2}} }} ,
\label{G}
\end{eqnarray}
where $\bm{\nu}$ and $\bm{u}$ are vectors with
$m=m_{\mathscr{V}}+m_{\mathscr{A}}$ components defined as
$\bm{\nu}^T = (\ldots \s{\nu}{V} \ldots; \ldots \s{\nu}{A} \ldots)$ and
$\bm{u}^T= (\ldots -\log[(x_0+V)/\epsilon] \ldots;\ldots \log A \ldots)$,
respectively.
For later use, we also define
$\bm{v}^T = (\ldots (x_0+V)^{-1} \ldots; \ldots 0 \ldots)$
and 
$\bm{w}^T = (\ldots (x_0+V)^{-2} \ldots; \ldots 0 \ldots)$.
Note that $\bm{u}$, $\bm{v}$ and $\bm{w}$ 
depend on $\bm{\nu}$ through $x_{0}=x_{0}(\bm{\nu})$ and
$\bm{v}=-\partial_{x_0}\bm{u}$, $\bm{w}=-\partial_{x_0}\bm{v}$.

The probability density $\mathcal{P}_N(\bm{\nu})$ is
given by the fraction of trajectories 
branching from the initial configuration $\bm{n}_0$ and having
after $N$ jumps multiplicities $N_V=\s{\nu}{V} N$ and $N_A=\s{\nu}{A} N$.
For $N$ large, it can be approximated in the following way. 
We rewrite the multiplicities as 
$N_V = \sum_{k=0}^N \s{\chi}{V}(\bm{n}_k)$ and 
$N_A = \sum_{k=1}^N \s{\chi}{A}(\bm{n}_{k-1})$,
where $\s{\chi}{V}(\bm{n})=1$ if $V(\bm{n})=V$ and $\s{\chi}{V}(\bm{n})=0$ 
otherwise, and similarly for $\s{\chi}{A}$.
Since the configurations $\bm{n}_k$ form a Markov chain
with finite state space, a central limit theorem applies to
each rescaled sum $N_V/\sqrt{N}$ or $N_A/\sqrt{N}$ \cite{BILLINGSLEY}.
Therefore, these rescaled variables are completely described in terms  
of the mean values $\overline{\bm{\nu}}\sqrt{N}$ and of 
the covariance matrix $\bm{\Sigma}$, which are easily 
measured by sampling over trajectories with a large number of jumps.
Due to the constraints (\ref{CONSTRAINTS}), it is easy to see that
\begin{eqnarray}
\label{MEANS}
\sum_{V\in\mathscr{V}} \s{\overline{\nu}}{V}=
\sum_{A\in\mathscr{A}} \s{\overline{\nu}}{A}=1, 
\end{eqnarray}
\begin{eqnarray}
\label{BLOCKS}
\sum_{V\in\mathscr{V}} {\Sigma}_{\alpha,V}=
\sum_{V\in\mathscr{V}} {\Sigma}_{V,\alpha}=
\sum_{A\in\mathscr{A}} {\Sigma}_{\alpha,A}=
\sum_{A\in\mathscr{A}} {\Sigma}_{A,\alpha}=0, 
\qquad \alpha \in \mathscr{A} \cup \mathscr{V}.
\end{eqnarray}

In the next two subsections we will describe two different 
choices for the density $\mathcal{P}_N(\bm{\nu})$.
The two densities differ only for the fact that the
first one satisfies the constraints (\ref{CONSTRAINTS}) in mean,
while the other one identically. 
We will show that, at least in the limit of large $N$,
the results obtained by assuming the two densities coincide.

\subsubsection{Purely Gaussian density}
In this case, we assume a purely Gaussian probability density 
\begin{eqnarray}
\label{GAUSSIAN}
\mathcal{P}_N({\bm{\nu}}) =
\sqrt{ 
\frac{N^{m} |\det {\bm{\Sigma}}^{-1}|} {(2\pi)^{m}} }
~e^{ -\frac{N}{2} 
\left( {\bm{\Sigma}}^{-1} ({\bm{\nu}} - 
{\overline{\bm{\nu}}}),
({\bm{\nu}} - {\overline{\bm{\nu}}}) \right) }.
\end{eqnarray}
Due to Eq. (\ref{BLOCKS}), we have that $\det \bm{\Sigma}=0$, 
\textit{i.e.} 
$\bm{\Sigma}$ is singular and the density (\ref{GAUSSIAN}) ill defined.
Nevertheless, this singularity is eliminable and 
the calculation with this density has the advantage to be rather simple.
We rewrite Eq. (\ref{AVERAGE1}) as
\begin{eqnarray}
\label{AVERAGE2}
\media{ \mathcal{W}_{N}(t) \! \prod_{A\in \mathscr{A}} A^{N_A} }  =
\int d\bm{\nu}
e^{N \phi(\bm{\nu})} R(\bm{\nu}), 
\end{eqnarray}
where 
\begin{eqnarray}
\label{PHI}
\phi(\bm{\nu})=
x_{0}\frac{t}{N}+ \left( \bm{\nu},\bm{u} \right) 
-\frac{1}{2}\left( {\bm{\Sigma}}^{-1} ({\bm{\nu}} - 
{\overline{\bm{\nu}}}),
({\bm{\nu}} - {\overline{\bm{\nu}}}) \right)
\end{eqnarray}
and $R(\bm{\nu})$ is a smooth function.
The parametric dependence of $\phi$ and $R$ on $t$ and $N$ 
is omitted for brevity.
We will perform the integral (\ref{AVERAGE2}) with a saddle-point method
by solving the $m$-dimensional stationary problem for $\phi(\bm{\nu})$.
The derivatives of $\phi$ with respect to the $V$ and $A$ 
components of $\bm{\nu}$ read
\begin{eqnarray}
\label{GRAD}
\left\{
\begin{array}{l}
\partial_{\s{\nu}{V}}\phi({\bm{\nu}}) =
\frac{t}{N} \partial_{\s{\nu}{V}}x_0 -
\left( {\bm{\nu}}, { \bm{v}} \right)
\partial_{\s{\nu}{V}}x_0 +  \s{u}{V}
-\left( {\bm{\Sigma}}^{-1} ({\bm{\nu}} - 
{\overline{\bm{\nu}}})\right)_V
\nonumber \\
\partial_{\s{\nu}{A}}\phi({\bm{\nu}}) =
\s{u}{A}
-\left( {\bm{\Sigma}}^{-1} ({\bm{\nu}} - 
{\overline{\bm{\nu}}})\right)_A .
\end{array}
\right.
\end{eqnarray}
On the other hand, by using the definition of $x_0(\bm{\nu})$, 
Eq. (\ref{X}), it is easy to see that
$\left( {\bm{\nu}}, { \bm{v}} \right)=t /N$, so that 
$\phi(\bm{\nu})$ is stationary for $\bm{\nu}=\bm{\nu}^\mathrm{sp}$ 
solution of the equation
\begin{eqnarray}
\label{saddle}
\bm{\nu}^\mathrm{sp} = \overline{\bm{\nu}} + 
\bm{\Sigma} \bm{u}(\bm{\nu}^\mathrm{sp}).
\end{eqnarray}
Note that, due to Eqs. (\ref{MEANS}) and (\ref{BLOCKS}), 
we have
$\sum_{V\in\mathscr{V}} \nu_V^\mathrm{sp}
=\sum_{A\in\mathscr{A}} \nu_A^\mathrm{sp}=1$
and only $m-2$ equations in (\ref{saddle}) are independent.
The result of the saddle-point integration then is 
\begin{eqnarray}
\label{AVERAGE3}
\media{ \mathcal{W}_{N}(t) \! \prod_{A\in \mathscr{A}} A^{N_A} } =
e^{N\phi(\bm{\nu}^\mathrm{sp})}R(\bm{\nu}^\mathrm{sp})
\sqrt{\frac{(2\pi)^m}{N^m|\det \nabla^2 \phi(\bm{\nu}^\mathrm{sp})|}}, 
\end{eqnarray}
where
\begin{eqnarray}
\label{PHISP}
\phi(\bm{{\nu}}^\mathrm{sp}) &=&
\left[ x_{0}\frac{t}{N}+ \left(\overline{\bm{\nu}},\bm{u}\right) + 
\frac12 \left( \bm{\Sigma} \bm{u},\bm{u}\right)
\right]_{\bm{{\nu}}=\bm{{\nu}}^\mathrm{sp}}
\end{eqnarray} 
and 
$\nabla^2 \phi(\bm{\nu}^\mathrm{sp})= 
- \bm{\Sigma}^{-1} - \bm{A}(\bm{\nu}^\mathrm{sp})$,
$\bm{A}$ being the $m \times m$ matrix with elements 
\begin{eqnarray}
{A}_{\alpha,\beta}=\frac{v_{\alpha}v_{\beta}}
{(\bm{\nu},\bm{w})},
\qquad 
\alpha,\beta \in \mathscr{V} \cup \mathscr{A}.
\end{eqnarray} 
By writing $\det \nabla^2 \phi(\bm{\nu}^\mathrm{sp})=
- \det \bm{\Sigma}^{-1} \det (\bm{1} + \bm{\Sigma}\bm{A})$
and observing that $\det \bm{\Sigma}^{-1}$ cancels out, 
Eq. (\ref{AVERAGE3}) becomes
\begin{eqnarray}
\label{AVERAGE4}
\media{ \mathcal{W}_{N}(t) \! \prod_{A\in \mathscr{A}} A^{N_A} } =
\frac{1}{\sqrt{\left| \det (\bm{1} + \bm{\Sigma}\bm{A} ) \right| }}
~\left.
\frac{e^{x_{0} t + N \left[ \left(\overline{\bm{\nu}}, \bm{u} \right) + 
\frac12 \left( \bm{\Sigma} \bm{u},\bm{u}\right) \right]}}
{\sqrt{2 \pi N \epsilon^2 (\bm{\nu},\bm{w})}}
\right|_{\bm{\nu}=\bm{\nu}^\mathrm{sp}}.
\end{eqnarray}
Note that for $\hat{V} \equiv 0$ the matrix $\bm{A}$ is uniform
and $\det (\bm{1} + \bm{\Sigma}\bm{A}) =1$.
In general, $\det (\bm{\Sigma}\bm{A})=0$ so that
$\det (\bm{1} + \bm{\Sigma}\bm{A})  \simeq 1$ up to 
terms of second order in $\bm{A}$.

\subsubsection{Density with constraints identically satisfied}
In the previous subsection we have evaluated the canonical
averages by exploiting a central limit theorem for
the rescaled variables $N_V/\sqrt{N}$ and $N_A/\sqrt{N}$.
However, due to the constraints (\ref{CONSTRAINTS}), 
the joint probability for these $m$ rescaled sums is not purely Gaussian. 
Given an arbitrary set of $m_{\mathscr{V}}-1$ 
$V$-like components and $m_{\mathscr{A}}-1$ $A$-like components, 
we can assume a joint probability density as the product 
of a Gaussian density for this set of $m-2$ variables
and two delta functions which take into account the constraints
(\ref{CONSTRAINTS}), \textit{i.e.}
\begin{eqnarray}
\label{DENSITY}
\mathcal{P}_N(\bm{\nu}) = \mathcal{F}_N({\hat{\bm{\nu}}}) 
~\delta \left( \sum_{V\in\mathscr{V}} \s{\nu}{V}-1 \right) 
~\delta \left( \sum_{A\in\mathscr{A}} \s{\nu}{A}-1 \right),
\end{eqnarray}
where $\mathcal{F}_N({\hat{\bm{\nu}}})$ is the normal density defined
in terms of the vector ${\hat{\bm{\nu}}}$ having the 
$m-2$ chosen components of $\bm{\nu}$
\begin{eqnarray}
\label{GAUSSIANC}
\mathcal{F}_N({\hat{\bm{\nu}}}) =
\sqrt{ 
\frac{N^{m-2} |\det {\hat{\bm{\Sigma}}}^{-1}|} {(2\pi)^{m-2}} }
~e^{ -\frac{N}{2} 
\left( {\hat{\bm{\Sigma}}}^{-1} ({\hat{\bm{\nu}}} - 
{\hat{\overline{\bm{\nu}}}}),
({\hat{\bm{\nu}}} - {\hat{\overline{\bm{\nu}}}}) \right) }.
\end{eqnarray}
In the following, with the symbol $\hat{}$ over a vector 
or a matrix we will indicate the projection onto the chosen
$m-2$ dimensional index space.

By using the density (\ref{DENSITY}), we rewrite Eq. (\ref{AVERAGE1}) as
\begin{eqnarray}
\label{AVERAGE3}
\media{ \mathcal{W}_{N}(t) \! \prod_{A\in \mathscr{A}} A^{N_A} }  =
\int d\hat{\bm{\nu}}
e^{N \hat{\phi}(\hat{\bm{\nu}})} \hat{R}(\hat{\bm{\nu}}), 
\end{eqnarray}
where $\hat{R}$ is a smooth function of $\hat{\bm{\nu}}$ and, 
if $V_*$ and $A_*$ are the two components chosen to reduce 
to $m-2$ the dimension of the index space,
\begin{eqnarray}
\label{phi}
\hat{\phi}({\hat{\bm{\nu}}}) &=&
\hat{x}_{0}\frac{t}{N}+ \left( {\hat{\bm{\nu}}},\hat{\bm{u}}(\hat{x}_{0}) 
\right)
+\left( 1 -\sum_{V\in\mathscr{V}\setminus V_*} \s{\hat{\nu}}{V} \right) 
\s{u}{V_*}(\hat{x}_{0})
+\left( 1 -\sum_{A \in \mathscr{A}\setminus A_*} \s{\hat{\nu}}{A} \right) 
\s{u}{A_*}
\nonumber \\ &&
-\frac{1}{2}\left( {\hat{\bm{\Sigma}}}^{-1} ({\hat{\bm{\nu}}} - 
{\hat{\overline{\bm{\nu}}}}),
({\hat{\bm{\nu}}} - {\hat{\overline{\bm{\nu}}}}) \right).
\end{eqnarray}
Note that $\hat{x}_0 = \hat{x}_0 ({\hat{\bm{\nu}}})$ 
is the solution of Eq. (\ref{X}) where 
the constraints (\ref{CONSTRAINTS}) have been explicited.
The function $\hat{\phi}$ can be rewritten as
\begin{eqnarray}
\label{phi1}
\hat{\phi}({\hat{\bm{\nu}}})=
\hat{x}_{0}\frac{t}{N}+ \left( {\hat{\bm{\nu}}},\hat{\delta\bm{u}} \right) 
-\frac{1}{2}\left( {\hat{\bm{\Sigma}}}^{-1} ({\hat{\bm{\nu}}} - 
{\hat{\overline{\bm{\nu}}}}),
({\hat{\bm{\nu}}} - {\hat{\overline{\bm{\nu}}}}) \right) +
\s{u}{V_*} + \s{u}{A_*},
\end{eqnarray} 
where $\delta \bm{u} = \bm{u} - \bm{u}_0$ and 
$\bm{u}_0^T = (\s{u}{V_*} \ldots \s{u}{V_*};\s{u}{A_*} \ldots \s{u}{A_*})$.
By evaluating the derivatives of $\hat{\phi}$ with respect to the components
of $\hat{\bm{\nu}}$, we get
\begin{eqnarray}
\label{GRADC}
\left\{
\begin{array}{l}
\partial_{\s{\hat{\nu}}{V}}\hat{\phi}({\hat{\bm{\nu}}}) =
\frac{t}{N} \partial_{\s{\hat\nu}{V}}\hat{x}_0 -
\left( {\hat{\bm{\nu}}}, \hat{\delta \bm{v}} \right)
\partial_{\s{\hat\nu}{V}}\hat{x}_0 + \s{\delta u}{V}
-\left( {\hat{\bm{\Sigma}}}^{-1} ({\hat{\bm{\nu}}} - 
{\hat{\overline{\bm{\nu}}}})\right)_V - \s{v}{V_*} 
\partial_{\s{\hat\nu}{V}}\hat{x}_0
\nonumber \\
\partial_{\s{\hat\nu}{A}}\hat{\phi}({\hat{\bm{\nu}}}) =
\s{\delta u}{A}
-\left( {\hat{\bm{\Sigma}}}^{-1} ({\hat{\bm{\nu}}} - 
{\hat{\overline{\bm{\nu}}}})\right)_A,
\end{array}
\right.
\end{eqnarray}
where  
$\delta \bm{v} = - \partial_{\hat{x}_0} \delta\bm{u} = 
- \left( \partial_{\hat{x}_0} \bm{u} - \partial_{\hat{x}_0} \bm{u}_0 \right)$.
Observing that 
$\left( {\hat{\bm{\nu}}}, \hat{\delta \bm{v}} \right) + \s{v}{V_*} = t/N$,
we can rewrite these derivatives as
\begin{eqnarray}
\label{GRADC2}
\left\{
\begin{array}{l}
\partial_{\s{\hat\nu}{V}}\hat{\phi}({\hat{\bm{\nu}}}) =
\s{\delta u}{V}
-\left( {\hat{\bm{\Sigma}}}^{-1} ({\hat{\bm{\nu}}} - 
{\hat{\overline{\bm{\nu}}}})\right)_V 
\nonumber \\
\partial_{\s{\hat\nu}{A}}\hat{\phi}({\hat{\bm{\nu}}}) =
\s{\delta u}{A}
-\left( {\hat{\bm{\Sigma}}}^{-1} ({\hat{\bm{\nu}}} - 
{\hat{\overline{\bm{\nu}}}})\right)_A.
\end{array}
\right.
\end{eqnarray}  
Therefore, the saddle-point equation is 
\begin{eqnarray}
\label{SPEQ}
\hat{\bm{\nu}}^\mathrm{sp} = \hat{\overline{\bm{\nu}}} + 
\hat{\bm{\Sigma}} \hat{\delta\bm{u}}(\hat{\bm{\nu}}^\mathrm{sp}).
\end{eqnarray}
Finally, we calculate $\hat{\phi}(\hat{\bm{\nu}}^\mathrm{sp})$. 
By using Eqs. (\ref{phi1}) and (\ref{SPEQ})
and the identities
$\left(\hat{\overline{\bm{\nu}}},\hat{\delta\bm{u}} \right)
+\s{u}{V_*}+\s{u}{A_*}
= \left(\overline{\bm{\nu}},\bm{u} \right) $
and 
$ \left( {\hat{\bm{\Sigma}}}
\hat{\delta \bm{u}},\hat{\delta \bm{u}}\right)=
\left( \bm{\Sigma} \bm{u},\bm{u}\right)$,
we get
\begin{eqnarray}
\label{phiSP}
\hat{\phi}(\hat{\bm{\nu}}^\mathrm{sp}) &=&
\left[ \hat{x}_0\frac{t}{N}+ 
\left(\hat{\overline{\bm{\nu}}} + 
\hat{\bm{\Sigma}} \hat{\delta \bm{u}}  
,\delta\hat{\bm{u}} \right) 
-\frac{1}{2}\left( {\hat{\bm{\Sigma}}} 
\hat{\delta \bm{u}},\hat{\delta \bm{u}}\right) 
+\s{u}{V_*}+\s{u}{A_*}
\right]_{\hat{\bm{\nu}}=\hat{\bm{\nu}}^\mathrm{sp}}
\nonumber \\ &=&
\left[ \hat{x}_0\frac{t}{N}+ 
\left(\overline{\bm{\nu}},\bm{u}(\hat{x}_0) \right) + 
\frac12 \left( \bm{\Sigma} \bm{u}(\hat{x}_0),\bm{u}(\hat{x}_0)\right)
\right]_{\hat{\bm{\nu}}=\hat{\bm{\nu}}^\mathrm{sp}}.
\end{eqnarray}
Due to the identity
$\left( {\hat{\bm{\Sigma}}}
\hat{\delta \bm{u}},\hat{\delta \bm{v}}\right)=
\left( \bm{\Sigma} \bm{u},\bm{v}\right)$,
we have $\hat{x}_0 (\hat{\bm{\nu}}^\mathrm{sp}) = 
x_0 (\bm{\nu}^\mathrm{sp})$, where $x_0 (\bm{\nu}^\mathrm{sp})$
is the value obtained with the purely Gaussian density.
Therefore, Eq. (\ref{phiSP}) shows that,
at least at the saddle-point level, the chosen probability density
provides the same results obtained in the previous subsection.

\subsection{Resumming the canonical series}
In order to evaluate the expectation 
$\E \left( {\cal M}^{t}_{\bm{n}_0} \right)$ 
we need to resum the series (\ref{EXPANSION}).
For $t\to \infty$, the sum can be substituted with the integral
\begin{equation}
\label{psiint}
\E \left( \mathcal{M}^t_{\bm{n}_0} \right) = 
\frac{1}{\sqrt{\left| \det (\bm{1} + \bm{\Sigma}\bm{A} ) \right| }}
\int dN 
\frac{e^{\psi(N)}}
{ \sqrt{2\pi N  \epsilon^2 
(\bm{\nu}^\mathrm{sp}, \bm{w}(x_0(\bm{\nu}^\mathrm{sp})))  }}, 
\end{equation} 
where
\begin{eqnarray}
\psi(N)= x_{0}(\bm{\nu}^\mathrm{sp})t + 
N \left[ 
(\overline{\bm{\nu}},\bm{u}(x_0(\bm{\nu}^\mathrm{sp}))) 
+ \frac12 (\bm{\Sigma} 
\bm{u}(x_0(\bm{\nu}^\mathrm{sp})),\bm{u}(x_0(\bm{\nu}^\mathrm{sp}))) 
\right].
\end{eqnarray}
The integrand in (\ref{psiint}) is exponentially peaked at 
$N=N^\mathrm{sp}$, where $N^\mathrm{sp}$ satisfies
$\partial_N \psi (N^\mathrm{sp})=0$.
For a generic $N$ we have 
\begin{eqnarray*}
\label{DPSI}
\partial_N \psi(N) &=& \left[ \left\{ t 
-N \left[ ( \overline{\bm{\nu}},\bm{v} ) +
\left( \bm{\Sigma} \bm{u}, \bm{v} \right) \right]
 \right\}
\partial_N x_{0} +
 (\overline{\bm{\nu}}{},{\bm{u}}) + \frac{1}{2} 
\left( \bm{\Sigma} {\bm{u}}, {\bm{u}} \right)
\right]_{\bm{\nu}=\bm{\nu}^\mathrm{sp}}
\nonumber \\ &=&
\left[ 
 (\overline{\bm{\nu}}{},{\bm{u}}) + \frac{1}{2} 
\left( \bm{\Sigma} {\bm{u}}, {\bm{u}} \right)
\right]_{\bm{\nu}=\bm{\nu}^\mathrm{sp}},
\end{eqnarray*}
where, due to Eqs. (\ref{X}) and (\ref{saddle}),
we noticed that the term 
$\left\{ t -N \left[ ( \overline{\bm{\nu}},\bm{v} ) +
\left( \bm{\Sigma} \bm{u}, \bm{v} \right) \right] \right\}$ 
vanishes for $\bm{\nu}=\bm{\nu}^\mathrm{sp}$.
The stationarity condition for $\psi(N)$ is, therefore,
\begin{eqnarray}
\label{E1}
\left[
(\overline{\bm{\nu}}{},{\bm{u}}) + \frac{1}{2} 
\left( \bm{\Sigma} {\bm{u}}, {\bm{u}} \right)
\right]_{\bm{\nu}=\bm{\nu}^\mathrm{sp},{N=N^\mathrm{sp}}}
=0.
\end{eqnarray}
Equation (\ref{E1}) is a time independent equation which determines
$\left. x_0 \right|_{\bm{\nu}=\bm{\nu}^\mathrm{sp},N=N^\mathrm{sp}}$
as a function of $\overline{\bm{\nu}}$ and  $\bm{\Sigma}$. 
According to Eq. (\ref{X}),
this means that the quantity $N^\mathrm{sp}$ increases 
linearly with $t$ so that 
$\left. x_0 \right|_{\bm{\nu}=\bm{\nu}^\mathrm{sp},N=N^\mathrm{sp}}$
becomes independent of time in the limit $t\to\infty$.
By evaluating the second derivative of $\psi(N)$
\begin{eqnarray}
\partial^2_N \psi (N) = 
- ( \bm{\nu}^\mathrm{sp}, \bm{v}(x_0(\bm{\nu}^\mathrm{sp})) )
\partial_N x_0(\bm{\nu}^\mathrm{sp}),
\end{eqnarray}
where 
\begin{eqnarray}
\partial_N x_0(\bm{\nu}^\mathrm{sp})  = \frac1N
~\frac{( \bm{\nu}^\mathrm{sp}, \bm{v}(x_0(\bm{\nu}^\mathrm{sp})) )}
{( \bm{\nu}^\mathrm{sp}, \bm{w}(x_0(\bm{\nu}^\mathrm{sp})) )
+ 
( \bm{v}(x_0(\bm{\nu}^\mathrm{sp})), 
\bm{\Sigma} \bm{v}(x_0(\bm{\nu}^\mathrm{sp})) )
},
\end{eqnarray}
and approximating $\psi(N) \simeq \psi(N^\mathrm{sp}) + \frac12 
\partial^2_N \psi (N^\mathrm{sp}) (N-N^\mathrm{sp})^2$ in the exponent
and $N \simeq N^\mathrm{sp}$ elsewhere, the integral (\ref{psiint})
reduces to a Gaussian one which gives
\begin{eqnarray}
\label{EMt}
\E \left( \mathcal{M}^t_{\bm{n}_0} \right) = 
\sqrt{
\frac{ 1 + \mathrm{tr} (\bm{\Sigma}\bm{A} )} 
{\left| \det (\bm{1} + \bm{\Sigma}\bm{A} ) \right| }}
\left. \frac{e^{x_0 t}}{ \epsilon (\bm{\nu}, \bm{v})}
\right|_{\bm{\nu}=\bm{\nu}^\mathrm{sp},N=N^\mathrm{sp}}.
\end{eqnarray} 
Note that, since the peak of the Gaussian at $N=N^\mathrm{sp}$ 
moves to infinity linearly with $t$  
while its width increases only as $\sqrt{t}$,
the result (\ref{EMt}) is asymptotically exact for $t\to \infty$.

According to Eq. (\ref{E0}), the result (\ref{EMt}) shows 
that the ground-state energy of the hard-core boson system is
\begin{eqnarray}
\label{E0B}
E_{0B} = 
- \left. x_0 \right|_{\bm{\nu}=\bm{\nu}^\mathrm{sp},N=N^\mathrm{sp}}.
\end{eqnarray}
Equation (\ref{E1}) is, therefore, the equation for the ground-state energy.
It defines $E_{0B}$ in terms of $\overline{\bm{\nu}}$
and $\bm{\Sigma}$ and explicitly reads 
\begin{eqnarray}
\label{E2}
0 &=&
-\sum_{V\in\mathscr{V}}
\s{\overline{\nu}}{V} \log\left(\frac{-E_{0B}+V}{\epsilon}\right)
+\sum_{A\in\mathscr{A}}
\s{\overline{\nu}}{A} \log\left(A\right)
\nonumber \\ &&
+\frac{1}{2}\sum_{V\in\mathscr{V}}\sum_{V'\in\mathscr{V}} {\Sigma}_{V,V'}
\log\left(\frac{-E_{0B}+V}{\epsilon}\right)
\log\left(\frac{-E_{0B}+V'}{\epsilon}\right)  
\nonumber \\ &&
-\sum_{V\in\mathscr{V}}\sum_{A\in\mathscr{A}} {\Sigma}_{V,A}
\log\left(\frac{-E_{0B}+V}{\epsilon}\right) \log\left(A\right)
\nonumber \\ &&
+\frac{1}{2}\sum_{A\in\mathscr{A}}\sum_{A'\in\mathscr{A}} {\Sigma}_{A,A'}
\log\left(A\right)\log\left(A'\right) .
\end{eqnarray}
In the case $\hat{V}\equiv 0$, the ground-state energy $E_{0B}^{(0)}$ 
can be solved analytically
\begin{eqnarray}
\label{E_{0B}0}
E_{0B}^{(0)}=-\epsilon
\exp\left[
\sum_{A \in\mathscr{A}} \s{\overline{\nu}}{A} \log(A)+
\frac{1}{2}\sum_{A\in\mathscr{A}} \sum_{A'\in\mathscr{A}} 
{\Sigma}_{A,A'} \log(A)\log(A') 
\right]. 
\end{eqnarray}

\section{Conclusions}
\label{conclusions}
By using saddle-point techniques and a central limit theorem, 
we have exploited an exact probabilistic representation of the
quantum dynamics in a lattice to derive analytical approximations 
for the matrix elements of the evolution operator of a system
of hard-core bosons in the limit of long times. 
The approach yields a simple scalar equation for the ground-state energy.
This equation depends on the values of the generalized potentials $V$
and of the kinetic quantities $A$, and on the statistical moments 
$\overline{\bm{\nu}}$ and $\bm{\Sigma}$ of their asymptotic 
multiplicities $N_V$ and $N_A$.
In turn, these moments depend only on the structure of the system
Hamiltonian, not on the values of the Hamiltonian parameters.
This implies that the statistical moments must be
measured \textit{una tantum} for a given Hamiltonian structure and, 
once $\overline{\bm{\nu}}$ and $\bm{\Sigma}$ are known, 
our approach provides the ground-state energy analytically 
as a function of the Hamiltonian parameters.

The ground state energies obtained with the present formula reveal
a small systematic error when compared with exact results \cite{OP1}.
In fact, a more accurate analysis which takes into account the large 
deviations neglected by the central limit theorem shows that
the present approach corresponds to a second order truncation 
of an exact cumulant expansion for the ground-state energy \cite{OP2}. 

Similar results hold in the case of Hamiltonians
with arbitrary kinetic operators.


\end{document}